\documentclass[preprint,preprintnumbers,amsmath,amssymb,superscriptaddress]{revtex4}
\usepackage{pslatex}
\usepackage{setspace}
\usepackage{amsmath}
\usepackage{amssymb}
\usepackage{graphicx}
\usepackage{epstopdf}
\usepackage{dcolumn}
\usepackage{bm}

\begin{document}

\title{Potential contributions of noncontact atomic force microscopy for the future Casimir force measurements}

\author{W. J. Kim}
\affiliation{Department of Physics, Seattle University\\ 901 12th Ave, Seattle, WA  98122}
\author{U. D. Schwarz}
\affiliation{Department of Mechanical Engineering and Center for Research on Interface Structures and Phenomena (CRISP) \\ Yale University, New Haven, CT  06511}

\date{\today}

\begin{abstract}
Surface electric noise, i.e., the non-uniform distribution of charges and potentials on a surface, poses a great experimental challenge in modern precision force measurements. Such a challenge is encountered in a number of different experimental circumstances. The scientists employing atomic force microscopy (AFM) have long focused their efforts to understand the surface-related noise issues via variants of AFM techniques, such as Kelvin probe force microscopy or electric force microscopy. Recently, the physicists investigating quantum vacuum fluctuation phenomena between two closely-spaced objects have also begun to collect experimental evidence indicating a presence of surface effects neglected in their previous analyses. It now appears that the two seemingly disparate science communities are encountering effects rooted in the same surface phenomena. In this report, we suggest specific experimental tasks to be performed in the near future that are crucial not only for fostering needed collaborations between the two communities, but also for providing valuable data on the surface effects in order to draw the most realistic conclusion about the actual contribution of the Casimir force (or van der Waals force) between a pair of real materials.

\end{abstract}

\maketitle

\section{Introduction}

At the 12th International Conference on Noncontact Atomic Force Microscopy (NC-AFM), which had recently been held on the campus of Yale University in New Haven, CT, USA, a special forum for bringing physicists applying NC-AFM together with their colleagues working in the field of Casimir effect had been instituted for the first time since the NC-AFM conference series was started in 1998. The two communities share an important common scientific agenda --- the investigation of fundamental electrodynamic interactions at the material interfaces. The operational principle of an atomic force microscope, which had been introduced 1986 by G. Binnig, C. F. Quate, and Ch. Gerber \cite{afm}, critically hinges upon the short-ranged electromagnetic interactions between the front atom of a sharp tip and individual atoms on a sample surface. Likewise, the Casimir physicists strive to understand the quantum vacuum effects, originating from similar electromagnetic fluctuations on a pair of macroscopic objects \cite{cas}. Similarly, the underlying theoretical description of the so-called Lifshitz theory \cite{lif} is frequently employed by both groups of experimenters. And yet, there has been no tangible effort thus far to bring together the two communities in a science meeting for discussions on the unified theme. This divergence, which grew rampantly over the past decade, is most likely to stem from a stark difference in attitude towards the core subject matter --- a microscopic, materialistic approach adopted by the NC-AFM community versus a macroscopic approach favored by the Casimir community, which is geared towards a verification of quantum vacuum phenomena that are referred to as the ``Casimir effect''.

From an historical point of view, the scientific theme explored by the two communities traces back to a number of landmark papers published in the 1950's and 1960's. These include the works by B. V. Derjaguin {\it et al.} \cite{der}, by W. Black {\it et al.} \cite{bla}, by D. Tabor and R. H. S. Winterton \cite{tabor}, and by those of J. N. Israelachvili \cite{isra1,isra2}, all of which report on experimental studies of molecular attraction such as van der Waals forces (both retarded and non-retarded) as well as surface forces. Particularly, the Casimir community has shown keen interest in one of the early experimental papers by M. J. Sparnaay in 1958 \cite{spar}, which was continuing scientific experiments on the subject of short-range attractive forces pioneered by B. V. Derjaguin {\it et al.}  \cite{der}. Today, this paper has found a special place in the Casimir literature as the first experimental attempt to the measure the quantum vacuum-induced effect.

More recently, the field of the Casimir force measurement has enjoyed an explosive growth in popularity after S. K. Lamoreaux published the results of his torsion-balance experiment in 1997 \cite{lam}, by which it established itself as a distinguished field of modern precision measurement \cite{bor}. This breakthrough was followed by subsequent reports of further experimental confirmation \cite{moh,cap,bres,decca}. The Casimir effect appears aesthetically appealing to many physicists and even to laymen, as it is based on a simple but elegant theoretical argument: the presence of virtual photons each carrying energy of $\frac{1}{2}\hbar\omega$ in space regulated by two electrically neutral objects leads to a mutual attraction between them. Whether the force is thereby to be attributed to the {\it fields} of electromagnetic fluctuations inside a boundary formed by two plates (as envisioned by Casimir) or to the dipole {\it sources} of electromagnetic fluctuation of quantum origins that constitutes the material boundary (as formualted by Lifshitz) is rather a topic of philosophical discussion \cite{mil}. Yet, for sure, this subtlety in the interpretation appeared to have created a certain dichotomy between the two communities while inducing the soaring popularity of the Casimir subject. With the remarkable 0.2~\% level of  agreement between theory and data in a recent Casimir force measurement \cite{decca2}, there has been a great deal of advancements in search for non-Newtonian gravitation interactions in the submicron regime with an implication for the existence of non-standard forces \cite{fuji,fish,long,stanford,ono,mas}.

Experimental efforts of the AFM community over the past decades have been mainly focused on practical issues surrounding force spectroscopy \cite{rugar,albre,gotsmann,hoelscher1,gug,hoelscher2,albers}. Significant developments have been made in precision electric force microscopy (EFM) \cite{col,gek,bon,luc}, Kelvin probe force microscopy (KPFM) \cite{kpm,jac,kol,lapp,gross}, and magnetic force microscopy (MFM) \cite{stipe} with a number of these papers devoted to improvements of force sensitivity via high $Q$-factor and high stability measurements at low temperature (see also the reviews in \cite{rev1,rev2,rev3} for further information). In the meantime, the AFM scientists have noticed unusual challenges in dealing with precise quantification of electric interaction between a tip and a sample in the presence of inhomogeneous surfaces \cite{col,bon}, the influence of residual charges and surface patch potentials \cite{nancy,ter,elad}, and stray electric fields \cite{stipe,cornell}. It should be noticed that most of these problems have been clearly pointed out  by the early short-range force measurements \cite{der,bla,tabor,isra1} as well as by the studies of contact potentials on metallic surfaces dating back to 1933 \cite{rose,revi} as a major experimental bottleneck when involving the real surfaces.

To say the least, the surface potential issue poses one of the greatest challenges in modern precision measurements that involve a pair of real metallic conductors. This is clearly evidenced by a number of recent large-scale projects like Gravity Probe- B (GP-B), Laser Interferometer Space Antenna (LISA), and Laser Interferometer Gravitational wave Observatory (LIGO) (see discussions below).  By organizing the Casimir 2009 Workshop as a satellite forum to the NC-AFM 2009 conference, we hoped to bring together the two science communities to stimulate discussion on this important experimental problem that seems now commonplace in other areas of physics. In the context of measuring a precision contribution of the Casimir force for a given sample, the following questions remain open and must be critically examined:  ``Could the surface effect ever be {\textit {accurately}} and {\textit{precisely}} taken into account in a precision force measurement? If so, how accurately can they be characterized to give a valid comparison with the Casimir-Lifshitz theory? Could one ever observe a {\textit {pure}} Casimir force in an experiment employing a pair of real samples that necessarily bear defects and inhomogeneity to some degrees?''. To answer these questions most effectively, we subsequently review experimental methodologies employed in force measurements as well as some of the calibration properties of such measurements.

\section{Experimental Methodology}
There are two main modes of operation in a force measurement using an AFM. The more straightforward one of the two is the so-called \emph{static mode} of operation, in which a deflection of a cantilever due to a force between a tip located at the very end of the cantilever and a sample is constantly monitored. The deflection is then directly proportional to the total force $F_{\rm{meas}}$ between the tip and the sample, which is expressed as
\begin{equation}
\label{eq1}
F_{\rm{meas}}(d,V)=F_{\rm{elect}}(d,V)+F_{\rm{fluct}}(d) \; .
\end{equation}
 $F_{\rm{elect}}$ is the electric force in response to an externally applied electric potentials $V$ at a certain separation gap distance $d$;  note that $F_{\rm{elect}}$ captures only the electric force that is responsive to the externally applied voltage and might not represent all electric forces existing in the system. Such forces can also be part of $F_{\rm{fluct}}$, which summarizes all forces that are a function of the distance only. It can be divided as follows:
\begin{equation}
F_{\rm{fluct}}(d)=F_{\rm{cas}}(d)+F_{\rm{surf}}(d)+F_{\rm{exotic}}(d) \; .
\label{eq2}
\end{equation}
Here, the first term $F_{\rm{cas}}$ reflects the Casimir force (or van der Waals interaction) and results from quantum and thermal fluctuations of electromagnetic fields in a vacuum regulated by the objects' boundaries. Its theoretical description is detailed in the Lifshitz formalism and can be attributed to fluctuation of charges and currents on the surface of the material boundaries.

The primary task for a Casimir experimentalist is to quantify the precise contribution of the first term $F_{\rm{cas}}$ in the presence $F_{\rm{surf}}$, which is caused by electric effects due to spatial fluctuations of surface potentials. The contribution of $F_{\rm{surf}}$ must be therefore distinguished from the usual electrostatic contribution $F_{\rm{elect}}$, which simply vanishes when a proper offset voltage is chosen. The last term $F_{\rm{exotic}}$ represents contributions from non-standard forces  such as non-Newtonian gravity, if it existed. Data analyses undertaken in all Casimir force measurements thus far have assumed (with the exception of \cite{ge,sven2}, though) that (a) $F_{\rm{elect}}$ is the only electric force present in the system, which can be compensated by a unique offset voltage in all explored gap separations, and that (b) there is no additional contribution to the force of electric origin that could be distance-dependent, simplifying Eq.\ (\ref{eq2}) to $F_{\rm{fluct}}(d)=F_{\rm{cas}}(d)+F_{\rm{exotic}}$. In what follows, we will show that these two assumptions are not necessarily true.

The electric force $F_{\rm{elect}}$ as it appears in Eq.\ (\ref{eq1}) exhibits a quadratic behavior with externally applied electric potentials. This quadratic relation facilitates an effective calibration of the system and identifies other distance-dependent contributions in the measured force
\begin{equation}
F_{\rm{meas}}(d,V)=k^{\rm{fr}}_{\rm{el}}(d)(V-V^{\rm{fr}}_{\rm{m}})^2+F_{\rm{fluct}}(d) \; .
\label{fpara}
\end{equation}
The first term in Eq.\ (\ref{fpara}) corresponds to $F_{\rm{elect}}$ and is minimized when $V=V^{\rm{fr}}_{\rm{m}}$. The voltage found this way is defined to be a contact potential difference (CPD) between the interfacing surfaces and can be extracted at different gap separations. Under the minimizing condition, the measured force is equal to a sum of all the distance-dependent interactions $F_{\rm{fluct}}(d)$. The system's response to the applied electric potentials is represented by the parabola curvature $k^{\rm{fr}}_{\rm{el}}$, whose values asymptotically grow as the tip-sample gap separation is reduced. The specific formula for $k^{\rm{fr}}_{\rm{el}}$ should be known {\it a priori} for exact characterization of its asymptotic behavior. All three parabola parameters obtained in a static mode, i.e., $k^{\rm{fr}}_{\rm{el}}$, $V^{\rm{fr}}_{\rm{m}}$, and $F_{\rm{fluct}}$, can be plotted against gap separations.

In the framework of noncontact atomic force microscopy, where the cantilever is oscillated in close proximity of the surface under investigation, the quantity of interest is the force gradient $G$. For oscillation amplitudes that are much smaller than the decay length of the forces investigated \cite{hoelscher1}, $G$ is directly proportional to the change in the square of the measured frequency, i.e., $G=\partial F/\partial d=\Delta\nu^2/(4\pi^2m_{\rm{eff}})$ with $m_{\rm{eff}}$ reflecting the effective mass of the oscillator and $\Delta\nu^2\equiv\nu^2_0-\nu^2_{\rm{meas}}$. $\nu_0$ is the eigenfrequency of the ``free'' oscillator far from the surface (i.e., $F_{\rm meas} = 0$) and $\nu_{\rm{meas}}$ represents the resonance frequency measured under the action of the force gradient. If the condition $\nu_0\approx\nu_{\rm{meas}}$ is met, then $\Delta\nu^2\approx2\nu_0\Delta\nu$, where $\Delta\nu$ is the change in frequency. This constitutes the second mode of AFM operation. Like in a static force measurement, different contributions to the dynamic force gradient can be identified by minimizing the electric force gradient at $V=V^{\rm{gr}}_{\rm{m}}$:
\begin{equation}
G_{\rm{meas}}(d,V)=k^{\rm{gr}}_{\rm{el}}(d)(V-V^{\rm{gr}}_{\rm{m}})^2+G_{\rm{dist}}(d) \; .
\label{gpara}
\end{equation}

The parabola curvature $k^{\rm{gr}}_{\rm{el}}$ is analogous to $k^{\rm{fr}}_{\rm{el}}$ and represents the system's response to the externally applied electric potentials. The difference is, however, that $k^{\rm{gr}}_{\rm{el}}$ as obtained from the force gradient measurement is proportional to the second derivative of the capacitance of the system, whereas the curvature from the force measurement $k^{\rm{fr}}_{\rm{el}}$ is proportional to the first derivative of that capacitance. Both of these curvature values grow asymptotically as the separation gap is reduced, with the gradient curvature displaying one-order higher dependence with distances. Most importantly, the parabola curvature enables to characterize the absolute gap separation as well as the calibration factor for a given measurement. Therefore, in the case of a sphere-plane configuration where the capacitance is given by $C(d)=2\pi R\epsilon_0\log(R/d)$ where $R(\gg d)$ is the radius of curvature for the sphere, we find
\begin{equation}
k^{\rm{fr}}_{\rm{el}}(d)=\frac{\alpha}{(d^{\rm{fr}}_0-d_r)^1}\propto\frac{\partial C}{\partial d}
\end{equation}
and
\begin{equation}
k^{\rm{gr}}_{\rm{el}}(d)=\frac{\beta}{(d^{\rm{gr}}_0-d_r)^2}\propto\frac{\partial^2 C}{\partial d^2} \; ,
\end{equation}
respectively. The absolute gap separation between the sphere and the plane is defined as $d\equiv d_0-d_r$, where $d_0$ is a fit parameter that provides the asymptotic limit of the given power law. $d_r$ is the relative distance recorded by an actuator during a measurement. For a static force measurement, $\alpha$ in $k^{\rm{fr}}_{\rm{el}}$ contains the calibration factor that converts the measured deflection values into the actual unit of force. For a force gradient measurement, $\beta$ in $k^{\rm{gr}}_{\rm{el}}$ comprises information about the effective mass of the cantilever. Mathematically, the gradient is simply related with a derivative of force and is proportional to the difference between the square of $\nu_0$ and the square of the measured frequency $\nu_{\rm{}meas}$ of the cantilever. Under ideal circumstances, the asymptotic limit given by each of the power laws must coincide with an actual physical barrier, such as surface roughness at a point of hard contact, i.e., $d_0^{\rm{fr}}=d_0^{\rm{gr}}$. In reality, however, one should not expect that the parabola variables including their curvature parameters as shown in Eqs.\ (5) and (6) necessarily provide a unique set of physical values. For instance, the CPD obtained from a static measurement $V^{\rm{fr}}_{\rm{m}}$ and its profile with respect to distances may look quite different from that obtained from a dynamic measurement $V^{\rm{gr}}_{\rm{m}}$, even if the identical samples are employed. Disparity could be seen through other parameters such as $d_0^{\rm{fr}}$ and $d_0^{\rm{gr}}$. This can affect the interpretation of the relevant contributions of different physical interactions to be sought in $F_{\rm{fluct}}$ and $G_{\rm{fluct}}$. In Table 1, we summarize various fit parameters involved in different modes of AFM operation including the mode operated for dissipation measurements in which a cantilever is made to oscillate on a plane defined by the surface of a sample, as shown in Fig.\ 1. The dissipation measurement, which is sometimes being referred to as ``friction'' measurement \cite{arv}, has occasionally also been used for the measurement of magnetic forces (see, e.g., Refs.\  \cite{stipe,cornell}), with its electrostatic calibration facilitating the quadratic relation with externally applied voltages shown in Eqs.\ (3)-(4).

\begin{table}[htbp]
\centering
\begin{tabular}{ @{} cccc @{} }
\hline
Method & Force (Static) & Gradient (Dynamic) & Dissipation (Friction) \\
\hline
\hline
Electric interaction & $F_{\rm{elect}}=k^{\rm{fr}}_{\rm{el}}(V^{\rm{fr}}_{\rm{m}}-V)^2$ & $G_{\rm{elect}}=k^{\rm{gr}}_{\rm{el}}(V^{\rm{gr}}_{\rm{m}}-V)^2$ & $D_{\rm{elect}}=k^{\rm{ds}}_{\rm{el}}(V^{\rm{ds}}_{\rm{m}}-V)^2$ \\
Parabola curvature & $k^{\rm{fr}}_{\rm{el}}=\alpha C'(d^{\rm{fr}}_0-d_r)$ &  $k^{\rm{gr}}_{\rm{el}}=\beta C''(d^{\rm{gr}}_0-d_r)$ & $k^{\rm{ds}}_{\rm{el}}=\gamma C^2(d^{\rm{ds}}_0-d_r)$\\
\hline
\end{tabular}
\caption{Equations describing electric interactions and their respective parabola curvatures in three distinct AFM operation modes:  (1) \emph{static mode}, in which a deflection of a cantilever is measured and related to the actual force $F$, (2) \emph{dynamic mode}, in which the frequency shift of the cantilever is monitored, yielding the gradient of the force interaction $G$, and (3) \emph{friction mode}, in which the quality factor of the cantilever resonance is monitored and used to quantify the overall dissipative interaction $D$. All three AFM modes reveal some aspects of the electromagnetic interactions between the tip and the sample, but they do not necessarily yield the same experimental information, such as characterizations of the distance-varying CPD and the asymptotic limits. Furthermore, each of the operation modes underscores a different physical mechanism of the interaction, making it difficult to compare among different sets of measurements.}
\end{table}

\begin{figure}[t]
\centering
\includegraphics[width=0.9\columnwidth,clip]{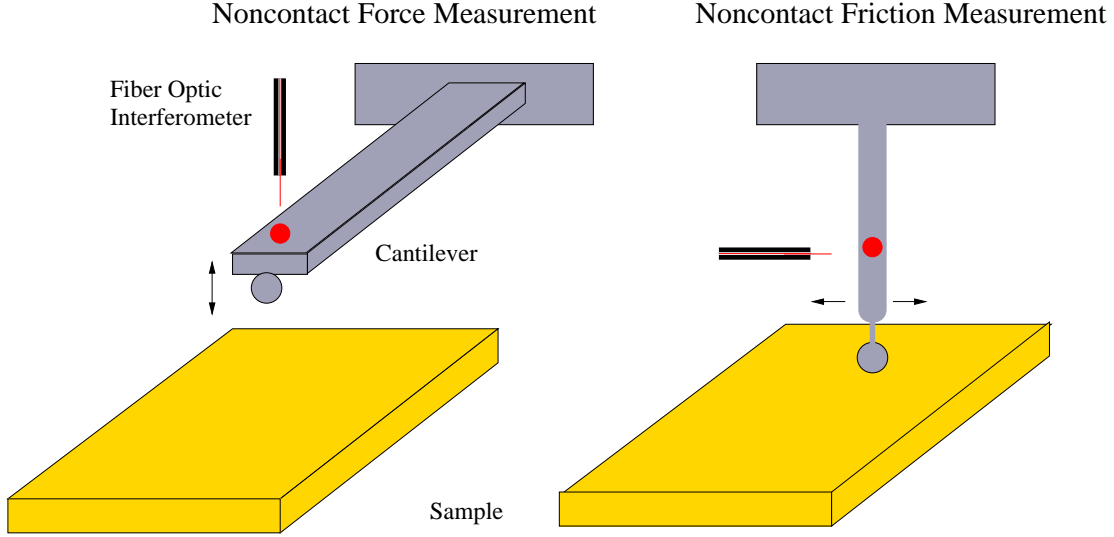}
\caption{Schematics of the experimental setup for both noncontact force (left) and noncontact friction measurements. The actuation of the cantilever in the force measurement is directed to be perpendicular to the plane defined by the sample surface in either static (force) or dynamic (gradient) modes. For the friction measurement, the cantilever is vertically situated with respect to the sample surface.}
\label{fig1}
\end{figure}

\section{Surface Electric Effects}
The key aspect of the {\it parabola} calibration applied in force measurements is to minimize the electrostatic interaction of the system and to extract only the distance-dependent term, which contains the possible contribution of quantum and thermal fluctuations, namely  $F_{\rm{fluct}}$, $G_{\rm{fluct}}$, or $D_{\rm{fluct}}$ (in the case of quantum friction measurement), respectively. These values are to be scrutinized to reveal the fundamental interaction of quantum and thermal origins, or other type of  surface fluctuation force. If one has an ideal system of perfectly conducting, smooth interfacing surfaces, there should exist an unique value of the CPD. This ensures that a single value of $V_{\rm{m}}$ will achieve the parabola minima in all gap separations, as shown in Fig.\ 2a. In this idealized situation operated in a static mode, the distance-dependent term $F_{\rm{fluct}}$ would be always electrostatic force-free, following the minimum interaction line of solid curve as shown in Fig.\ 2c. Therefore, as long as the CPD remains constant (e.g. $V^{\rm{fr}}_{\rm{m}}(d)\equiv V^{\rm{fr}}_{\rm{m}}$), the experimental procedure is relatively simplified.
\begin{figure}[t]
\centering
\includegraphics[width=0.9\columnwidth,clip]{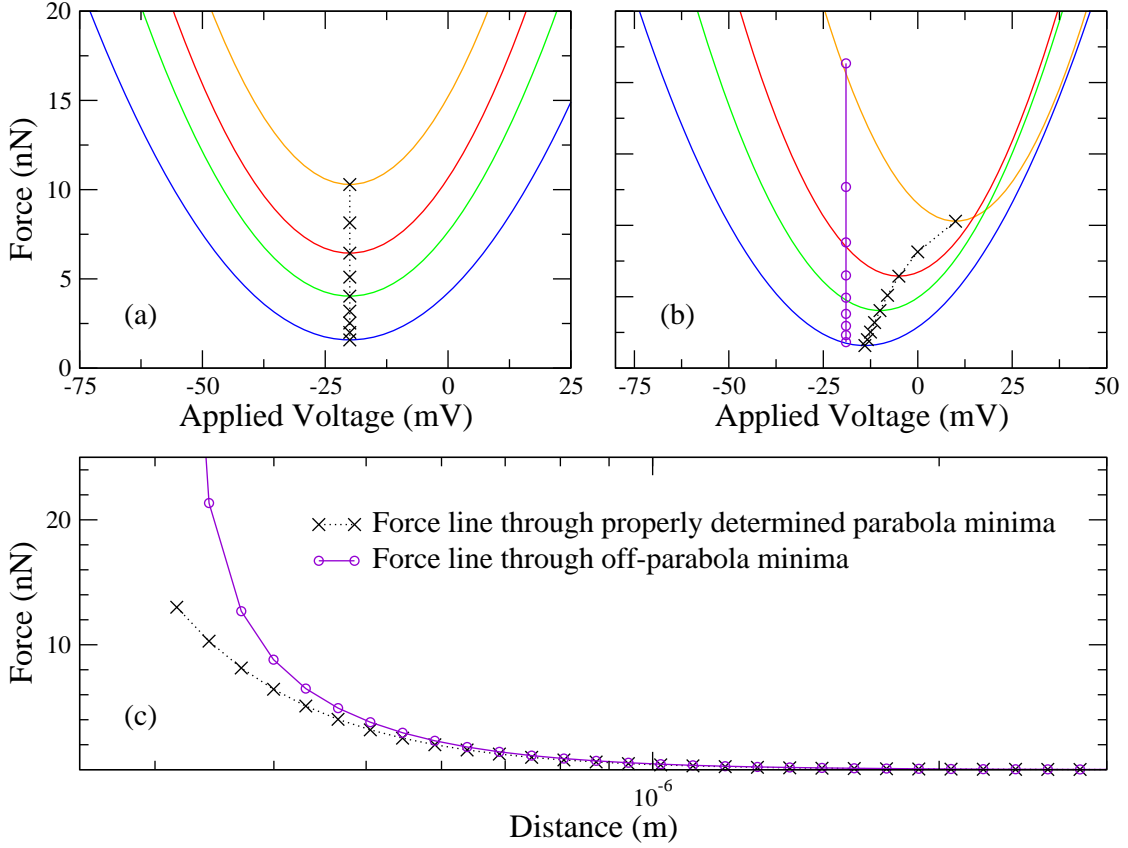}
\caption{The forces acting on the cantilever as a function of the applied voltage for the ideal case (a), where a single value of $V_{\rm m}$ describes the parabola minima at all distances, and for the non-ideal case of a distance-dependent $V_{\rm m}$ (b). Parabola measurements showing a condition of properly and improperly determined energy minima are indicated by dotted and solid lines, respectively. The bottom curve illustrates the degree of overestimation of the force curve as a result of an ``off the path" condition away from the proper minima.}
\label{fig2}
\end{figure}

If, however, the CPD varies with distance, one must continue to perform parabola measurements to follow the correct {\it path} given by the minimum condition for electric energy. This path is marked by a dotted line following cross marks in Fig.\ 2b. Inadvertent application of a constant value of the CPD in this case will cause a significant deviation from true minima of the electric energy, resulting in an overestimation of the distance-dependent interactions, as shown by a solid line running over circle marks in Figs.\ 2b and 2c. Recent precision force measurements indeed report that the CPD varies with gap separations \cite{kim,sven,kim2,gund,ge}, even in a situation where the system faithfully obeys the elementary electrostatic power law given by $k_{\rm{el}}$ \cite{sven}. Note that $k_{\rm{el}}$ represents the system's response to externally applied voltages. Its adherence to the basic Coulomb's electrostatic law, as far as implications given by the recent measurements are concerned, does not appear to be affected by the presence of the distance-dependent $V_{\rm{m}}$, although Kim {\it et al.} \cite{kim} found not only distance-dependent values of $V^{\rm{gr}}_{\rm{m}}$ , but also $k^{\rm{gr}}_{\rm{el}}$ values that reveal an anomalous scaling in the basic Coulomb interaction. Similar findings of distance-dependent CPD values have been reported independently from the AFM community \cite{kol,gek,boc}, ascertaining the recent experimental observations made by the Casimir community.

\section{Relevance in other areas of physics}
One possible explanation for the observation of distance-dependent CPD values is the presence of patch charges \cite{kim,ge} caused by an inhomogeneous, random distribution of surface defects such as atomic steps and pits, adsorbates, strains, impurities, and different crystallographic orientations. Such defects are unavoidable, regardless of how careful a sample surface has been prepared, a fact which ultimately sets a fundamental limit to precision force measurements between two closely spaced bodies \cite{stipe,speake}. These defects are widely encountered across a broad range of subfields in physics. Hence, dedicated studies of the surface potential effect to reveal the origins of the distant-dependent CPD would be highly desirable and will make a valuable contribution to many areas in physics. Here, we describe a few important experiments that have been negatively affected by the inevitable presence of the surface effects.

{\it Friction measurements:} Stipe {\it et al.\/} report the measurement of dissipation (``noncontact friction'') between a gold tip and a gold sample using an AFM operating in a configuration similar to the one depicted in Fig.\ 1b \cite{stipe}. The friction experienced by the tip of the cantilever is measured by monitoring the linewidth of its oscillation resonance. Thereby, significant dissipation is found and attributed to an inhomogeneous distribution of stray electric fields, which is confirmed by a subsequent test using fused silicon samples in which the dissipation associated with the surface electric effect was far greater pronounced. Another report employing a similar geometric setup, but with a magnetically sensitive tip over a magnetic sample concludes dielectric fluctuations that produces time-varying electric fields (at the capacitively charge tip) as the origin of noncontact friction \cite{cornell}. Results of all these experiments strongly indicate a situation in which the friction due to the surface electric fields overwhelms the fluctuations of quantum and thermal origins such that $D_{\rm{surf}}\gg D_{\rm{cas}}$. Note that in a friction measurement the electric term $D_{\rm{elect}}$ is also quadratic with respect to applied electric potentials and the curvature of such a parabola curve is proportional to the square of the capacitance of the interfacing materials \cite{cornell}.

{\it Force measurements:} Burnham {\it et al.} notice a significant deviation of the measured force from the value that would be theoretically expected \cite{nancy} in a van der Waals force measurement, with a long-ranged electric force dominating the short-ranged dispersive force. Examining the anomaly by a model of anisotropic distributions of patch charges between the two interacting surfaces, the authors conclude ``Recognizing the existence of patch charge force has many consequences. It strongly suggests that previous interpretations of force curve data are incomplete'', accentuating the preponderant presence of the surface electric force in fundamental studies of electromagnetic interactions. More recently, Speake and Trenkel have considered an electric force due to spatial variations of surface patch potentials and show that this force could in principle completely conceal the Casimir force \cite{speake}, reaching a similar conclusion made in \cite{stipe}: $F_{\rm{surf}}\gg F_{\rm{cas}}$, even when $F_{\rm{elect}}$ is minimized at proper CPDs in all gap separations. The theoretical prediction of the surface potential fluctuations by Speake and Trenkel, together with a long-range electric force in the measured data as noticed by Burham {\it et al.}, have been experimentally observed with a torsion balance setup \cite{ge}. All these findings once again indicate the need of detailed investigation of the surface electric force in real materials employed in a force measurement.

{\it Detection of gravitational waves and test of general relativity:} The surface potential effect transcends the domain of the force and friction measurements. Its presence is constantly recognized in fundamental research on gravitational waves, such as in the Laser Interferometer Gravitational Wave Observatory (LIGO) \cite{ligo,har}, the Laser Interferometer Space Antenna (LISA) \cite{speakelisa,lisa,lisa2,gund,vitale}, as well as in the mission of Gravity Probe B (GP-B) \cite{lisa}. One technical limitation for LIGO is the accumulation and motion of surface charges between the interferometer cavity mirror and the suspension frame. These charges will eventually generate static electric fields that induce a force, introducing additional noise to positioning accuracy. Likewise, the surface potential effect is one of the largest noise sources for LISA. With its stringent temporal stability requirement ($10^{-12}$ m resolution maintained over a period of 1000 seconds), the sensitivity of LISA can be easily destroyed by a presence of force larger than $10^{-16}$ N on its proof mass. This implies that fluctuation in electric potential between the proof mass and a set of electrodes surrounding it cannot exceed 0.01 mV. Given that a typical contact potential measurement yields a variation of 1 mV or larger between two metallic surfaces \cite{camp1,camp2}, the surface fluctuation becomes an immediate bottleneck for the LISA mission. Furthermore, the data analysis with the GP-B mission launched in 2004 clearly reveals that the underlying causes of the unexpected energy dissipation on the motion of polhode is the interaction between the patch fields in the gyro rotor and the surfaces of the inside housing. The damping of the gyro rotor caused by the surface patch effect being the major systematic effect, the status of one of its main missions, Frame-Dragging Effect, is still inconclusive.

{\it Quantum Point Contact:} Issues related to the surface effect are conspicuous in condensed matter physics, particularly in the subject of quantum point contacts (QPC). Conductance quantization is a well-known physical phenomenon leading to the universal conductance of $G_0=2e^2/h$, first observed in a two-dimensional electron gas semiconductor structure \cite{2dg} and later demonstrated using macroscopic metallic wires by table-top experiments based on scanning tunneling microscopy \cite{kram}. When two wires are brought in and out of contact at a fixed voltage $V$ between them, the quantized conductance appears as a result of transient formation of a single nanowire bridge at the wire juncture. The electric currents $I_{\rm{m}}$ are then measured right before rupture and the conductance value is obtained from the Ohm's law $I=V/R_0$, where $R_0\equiv1/G_0$. It turns out that the measured conductance (resistance) is always smaller (larger) than the expected $G_{\rm{m}}<G_0$ ($R_{\rm{m}}> R_0$) with primary suspects being impurities and disorder present inside nanowires \cite{diso}. This implies that the measured current $I_{\rm{m}}$ is underestimated because of the larger-than-expected resistance. This is equivalent to stating that the voltage applied at the junction is smaller, possibly due to a presence of surface contact potential difference $\Delta V_{\rm{eff}}$, while the resistance (conductance) remains to be $R_0$ ($G_{0}$). Hence, the residual resistance effectively translates to the effective contact potential $\Delta V_{\rm{eff}}=V(1-R_0/(R_0+R_{\rm{r}}))$, which yields $\Delta V_{\rm{eff}}\approx 1$ mV when the applied junction voltage is 50 mV and the measured residual resistance is $R_{\rm{r}}\approx 400$ $\Omega$ \cite{diso}.

{\it Freely falling electrons under gravity:} In the late 1960s, motivated by various theoretical predictions regarding gravitational effects on particles and antiparticles, measurements of the gravitational force on freely falling electrons were performed inside a metallic enclosure made of copper \cite{fair,fair2}. Briefly, these experiments employ the time-of-flight method by directly measuring the flight time of freely falling electrons. Not so surprisingly, the principal difficulty of the experimental analysis was how to account for spatial variations in the work function at the surface of the metallic enclosure, which induce additional electric fields interacting with the falling electrons. A discrepancy existed between the results of the early measurement \cite{fair}, with the expected magnitude of the surface fields at room temperature being larger by four orders of magnitude. Later, it was demonstrated that these ambient fields drop sharply at a temperature of 4.2 K at which previous measurements had been performed \cite{fair3}. Although the apparent discrepancy is resolved in this particular context, the main result regarding the temperature-dependent surface shielding effect on metals presents a challenge of its own, especially when considering most experiments testing two-body interaction in close proximity up to date have been carried out at a temperature much higher than 4.2K.

{\it Neutral atoms and ion trap experiments:} Electric field noise has become one of the major problems in the studies of neutral atoms \cite{suk,eric} and ion traps \cite{mich,lab1}. A detailed investigation of surface electric field noise above gold surfaces has been performed by a group at MIT who has observed a strong dependence of the noise on temperature (7 to 100 K) and frequency (0.6 to 1.5 MHz) \cite{lab2}. Their findings are consistent with the noncontact friction measurements in \cite{cornell} discussed above, suggesting the common origin of fluctuating patch potentials inevitably present on metallic surfaces. An AFM study of the CPD dependance on temperature has been also investigated in \cite{temp}.

\section{Conclusion}
Surface electric noise is pervasively present in many of the modern setups used for performing high-precision force measurements, posing as a unique experimental challenge that requires broader collaborations across the field. Clearly, an AFM seems the most suitable choice to investigate the underlying problem as it can collect \emph{local} information about the surface electrical properties. In order to make the most reasonable assessment about the actual contribution of the Casimir force (when interfered by the presence of the unknown surface electric forces), we propose the following experimental activities to be performed by both the AFM specialists and the Casimir experimenters in the near future.
\begin{itemize}
\item{\emph{Origins of the CPD and its distance-dependency:} Many of the recent Casimir force measurements have shown the distance-varying contact potential and its possible role in the long-range electric force $F_{\rm{surf}}$ that is to be distinguished from the contribution from the Casimir force $F_{\rm{cas}}$ \cite{kim,sven,kim2,ge,ge2}. Note that a systematic study of the noise issue faced in the LISA project has also shown the distance-varying CPD \cite{gund}, while the AFM community has noticed the behavior in different experimental contexts \cite{gek,kol}. Why does the CPD depend on gap separations? Can it be avoided with a careful preparation of a sample? Dedicated studies on this issue will be extremely valuable.}
\item {\emph{Comparison of parabola parameters obtained from different operation modes:} Can the CPD measurement be taken to be unique for a given sample? Or is this a by-product of a specific mode of AFM operation? Put another way, would $V^{\rm{fr}}_{\rm{m}}$, $V^{\rm{gr}}_{\rm{m}}$, and $V^{\rm{ds}}_{\rm{m}}$ be the same and obey the same distance-dependency? The distance-dependency due to the sample surface is one thing, but its mutability with respect to a particular mode of operation is another issue. Likewise, verifying  whether the parabola curvatures $k^{\rm{fr}}_{\rm{el}}$, $k^{\rm{gr}}_{\rm{el}}$, and $k^{\rm{ds}}_{\rm{el}}$ obey the elementary power laws indicated in Table 1 is an important systematic check. One should also make sure that the asymptotic limits given by the different modes of operation yield a common point of hard contact.}
\item{\emph{Identification of specific physical origins for the distance-dependent forces:} The central task in a precision force measurement is to identify the exact contribution of a particular origin of force. Even at the minimized condition, the electric force is never completely nullified and makes an additional contribution in the measured force through either $F_{\rm{fluct}}$, $G_{\rm{fluct}}$, or $D_{\rm{fluct}}$. Then, how precisely and accurately could one quantify the relevant contributions in the distance-dependent forces? Would there be any correlation between the CPD profiles and their corresponding distance-dependent surface forces $F_{\rm{surf}}$, $G_{\rm{surf}}$, and $D_{\rm{surf}}$? This issue has been recently addressed in some detail \cite{ge2}, but further experimental activities in this direction would be valuable.}
\item{\emph{Providing extensive data for different geometrical scales:} In most surface characterizing experiments using an AFM, the involved tip often probes local interactions between the tip and the sample. Exploring this boundary is somewhat tricky, and one could even misinterpret the usual local interaction between atoms for a macroscopic interaction (of quantum origin) between the bulk samples employed. No clear-cut procedure exists yet to distinguish one force from another. More analysis on what is actually being measured is needed. We hope that the activities proposed here will make unique contributions to our current understanding of electromagnetic interactions between real materials. They will benefit not only the Casimir  and AFM communities through extensive cross-talks and collaborations, but also experimenters in other areas in physics such as, in particular, in the field of precision measurements, who face the similar experimental challenge.}
\end{itemize}

{\small To whom correspondence should be addressed: kimw@seattleu.edu}

\begin{acknowledgments}
The authors are grateful for the generous support by Yale University for the NC-AFM 2009 Conference and Casimir 2009 Workshop, which took place August 10-14, 2009, on Yale campus, as well as by the European Science Foundation through its ``New Trends and Applications of the Casimir Effect'' program (PESC-2483). W. J. Kim acknowledges stimulating conversations with the Casimir experimenters of the third generation over the past two years: G. Jourdan, J. Munday, S. de Man, P. J. Van Zwol, G. Torricelli, Q. Wei, M. Brown-Hayes, and A. Sushkov. He also thanks S. K. Lamoreaux, R. Onofrio, and M. Bassan for many great discussions. 
\end{acknowledgments}

\end{document}